\documentclass[aps,twocolumn,superscriptaddress]{revtex4-2}
\usepackage{amsmath}
\usepackage{amsfonts}
\usepackage{amssymb}
\usepackage{graphicx}
\usepackage{xcolor}
\usepackage{gensymb}
\usepackage{siunitx}
\usepackage[colorlinks,citecolor=green,urlcolor=blue,bookmarks=false,hypertexnames=true]{hyperref} 
\usepackage{physics}
\usepackage{float}
\usepackage[export]{adjustbox}

\newcommand{\bk}{\vb{k}}

\begin{document}

\title{Exploring Grassmann manifolds in topological systems via quantum distance}

\author{Shin-Ming Huang}
 \affiliation{Department of Physics, National Sun Yat-sen University, Kaohsiung 80424, Taiwan} 
\affiliation{Center for Theoretical and Computational Physics, National Sun Yat-sen University, Kaohsiung 80424, Taiwan}
\affiliation{Physics Division, National Center for Theoretical Sciences, Taipei 10617, Taiwan}
\author{Dimitrios Giataganas}
\affiliation{Department of Physics, National Sun Yat-sen University, Kaohsiung 80424, Taiwan}
\affiliation{Center for Theoretical and Computational Physics, National Sun Yat-sen University, Kaohsiung 80424, Taiwan}
\affiliation{Physics Division, National Center for Theoretical Sciences, Taipei 10617, Taiwan, Republic of China}

\date{\today}

\begin{abstract}

Quantum states defined over a parameter space form a Grassmann manifold. To capture the geometry of the associated gauge structure, gauge-invariant quantities are essential. We employ the projector of a multilevel system to quantify the quantum distance between states. Using the multidimensional scaling method, we transform the quantum distance into a reconstructed manifold embedded in Euclidean space. This approach is demonstrated with examples of topological systems, showcasing their topological features within these manifolds. Our method provides a comprehensive view of the manifold, rather than focusing on local properties.

\end{abstract}

\maketitle

\section{Introduction}
Since the recognition of the Berry phase \cite{berry1984,Simon1983, Aharonov1987, Zak1989,resta2011} and the demonstration of topology in quantum Hall systems \cite{TKNN,Avron1983,kohmoto1985}, a new understanding of quantum physics has been developing. The electronic band structure and certain quantum numbers are not the only features that reflect physical properties. Eigenstates in some parameter spaces, such as momentum space, develop an abstract geometric structure in addition to providing transition probabilities \cite{Torma2023}. The Berry phase is a holonomy that manifests the geometric curvature in a projected Hilbert space, and its classification around non-contractible loops reveals the topological properties of the geometry.

The topological classification of Hamiltonians in different symmetry classes has been comprehensively accomplished in insulators, superconductors, and even semimetals, where states are characterized by corresponding topological invariants \cite{schnyder2008,kitaev2009,slager2013space,chiu2016,Bansil2016}. A key feature of a topological system is the bulk-boundary correspondence \cite{Laughlin1981, Hatsugai1993, Hasan2010, Essin2011}, which is the close relationship between the topological invariant and the appearance of edges or surface states within the band gap. This indicates the impossibility of isolating the valence bands from the conduction bands due to the obstruction of deforming states into exponentially localized Wannier functions \cite{brouder2007,bradlyn2017}. Consequently, long-range entanglement is a generic feature in topological quantum systems. The entanglement entropy and spectrum have become methods to detect topological order \cite{Levin2006,Kitaev2006,Ryu2006,Li2008,Pollmann2010,Turner2010,Hughes2011}.

Recently, the study of local quantum geometry has intensified. In the condensed matter field, ``local" refers to a crystal momentum (momentum, in short). This is captured by the quantum geometric tensor over momentum space \cite{provost1980,kolodrubetz2017,Torma2023}, a gauge-invariant complex matrix. The real part is the (Fubini-Study) quantum metric, and the imaginary part is half of the Berry curvature of the periodic Bloch wavefunction. The quantum metric measures the distance between quantum states in nearby momenta, while the Berry curvature provides information about the phase change of the eigenstate through a tiny round trip. Although they account for different information, they are closely related.

Remarkably, an inequality holds between the quantum metric and the Berry curvature \cite{roy2014,ozawa2021}. Integrating over momentum space, this inequality implies that the quantum volume (size of the manifold) becomes the upper bound of the absolute value of the topological invariant \cite{mera2022}. This rule is a good indication of nontrivial topology; in some models, the two quantities have an exact relation and have been measured in experiments \cite{Palumbo2018,Yu2019,Tan2021,gianfrate2020}.

Although the geometry of the manifold is abstract, the trace of the quantum metric is related to the spatial spread of the Wannier function \cite{marzari1997} and accordingly connects to the localization theory of insulators \cite{Kohn1964,Resta1999,Souza2000,hetenyi2013,Matsuura2010}. The geometric interpretation of the superfluid weight has also been explored, which is helpful for understanding high-$T_c$ superconductivity \cite{Peotta2015,Liang2017,Xie2020,HA2022,daido2024,hu2024quantum}. The investigation of geometric structures via optical responses is therefore intensively progressing, promising nonlinear optical applications \cite{neupert2013,xiang2024,Bac2024,ahn2020,ahn2022,Onishi2024Bound,Onishi2024structure}. Additionally, a theory on the connection between quantum geometry and electron-phonon coupling has been proposed \cite{yu2024}.

The local geometry is only part of the story in physics. A theoretical study has uncovered a universal behavior: the total Landau level spread is determined by the maximal quantum distance among single-particle states in the absence of the magnetic field \cite{rhim2020}. Intriguingly, the maximum quantum distance also influences the transport scattering rate, suggesting new directions for thermoelectric device research \cite{Oh2024}. Very recently, the concept of quantum geometric nesting has been proposed to address the interplay between quantum distance and interaction in a flat band system \cite{han2024}. Thus, understanding the entire manifold geometry is crucial for comprehending quantum systems. Furthermore, the exploration of manifolds is a significant topic in data science, particularly in optimization problems and machine learning \cite{Absil_book,hu2020brief,hsu2024quantum} and machine learning \cite{zhang2018grassmannian,wang2019unsupervised}.

In this paper, we present a comprehensive explanation of Grassmann manifolds (or Grassmannians) of non-interacting topological systems. A Grassmannian is a generalization of the projective space. It is a coset space by identifying gauge symmetry.  In terms of projectors, the Grassmannian is embedded in an Euclidean space $\mathbb{R}^n$ with a distance metric. We evaluate the quantum distance between two many-level quantum states and explain their geometry picture and gauge structure. By calculating the distances among eigenstates and utilizing the classical multidimensional scaling technique, the abstract manifold of a quantum state is embedded in an Euclidean space. With this embedding, we can visualize the structure of a manifold. The nontrivial topology is also evident from the appearance of the ``hole" in the manifold. 

The paper is organized as follows. In Sec.~\ref{sec:qd}, we explain the quantum distance and its geometry idea. In Sec.~\ref{sec:results}, we exemplify some non-interacting topological systems by showing their quantum distance map and manifold geometry over momentum space: 2D Chern inuslators, including two-band and four-band systems in Sec.~\ref{sec:2D_Chern}, a 2D time-reversal topological inuslator in Sec. \ref{sec:2D_TI}, a 3D Hopf insulator in Sec. \ref{sec:hopf}, and a 3D axion insulator in Sec. \ref{sec:axion}. Lastly, we conclude the paper in Sec. \ref{sec:conclusion}.

\section{Quantum distance} \label{sec:qd}
Now we are going to define the distance between the two systems. At first, we illustrate the distance between two states $\ket{u}$ and $\ket{v}$. Since $\ket{u}$ and $a \ket{u}$, $a \in \mathbb{C}$, stand for an equivalent state, a meaningful quantity is the following:
\begin{equation}
    \cos \theta_{u,v} = \frac{\abs{\braket{v}{u}}}{\norm{u} \norm{v}},
\end{equation}
where $\theta_{u,v}$ is named the principal angle, ranging between $0$ and $\pi/2$. We define the \emph{quantum disparity} of states as
\begin{equation} \label{duv}
    d_{u,v} = \frac{2}{\pi} \theta_{u,v}.
\end{equation}
to limit $ d_{u,v} \in [0, 1]$. When $\ket{u}$ and $\ket{v}$ are the same state, $ d_{u,v}=0$; when $\ket{u}$ and $\ket{v}$ are orthogonal to each other, $ d_{u,v}=1$.

Considering a system with $n$ energy levels, where $k$ and $n-k$ are the number of filled and unfilled states respectively, the collection of the filled states is represented by the matrix $\Psi = \left\{ \ket{\psi_1},\ket{\psi_2},\hdots,\ket{\psi_k} \right\}$, where each $\ket{\psi}$ is an $n$-component complex vector orthogonal to the others. Assuming the kets are normalized, $\Psi$ is interpreted as a sub-coordinate in $n$ dimensions. The system remains physically invariant when the basis (or sub-coordinates) of either filled or unfilled states is rotated by a unitary transformation. Thus we consider the system as a $k$-plane ($k$-dimensional plane) through the origin. 
Mathematically, $\Psi$ belongs to the complex Grassmann manifold $G_{k,n}(\mathbb{C}) \cong \text{U}(n)/\left(\text{U}(k)\times \text{U}(n-k) \right)$, which is the span of $n \times k$ complex matrices of rank $k$ \cite{nakahara2018geometry}. $\Psi$ and $\Psi'$ are considered the same point in $G_{k,n}(\mathbb{C})$ when $\Psi' = \Psi U$ for any unitary matrix $U \in \text{U}(k)$. Therefore, we sometimes use $\Psi$ to represent a point in the Grassmann manifold rather than a specific matrix. A point in the manifold corresponds to a $k$-plane. The Plücker embedding is introduced by Bouhon et al.~\cite{bouhon2023quantum} by identifying the multilevel state as a single vector in a $\binom{n}{k}$-dimensional complex Hilbert space whose basis are wedge products of single-level basis vectors. 

\begin{figure}[tbp]
    \centering
    \includegraphics[width=0.45\textwidth]{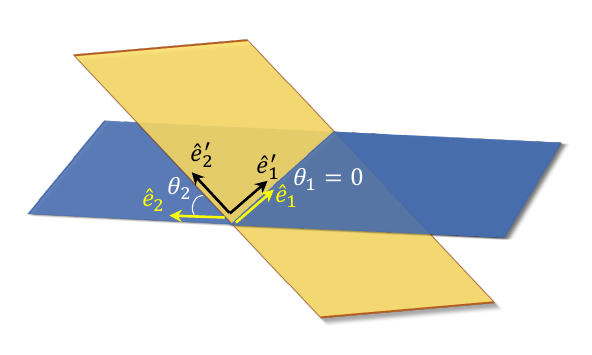}
    \caption{Principal angles $\theta_1$ and $\theta_2$ of two nonparallel planes. The intersection line of two planes determines the first basis $\hat{e}_1$ and $\hat{e}_1^{\prime}$ in two planes, giving $\theta_1=0$. The second principal angles $\theta_2$ is the relative angle between the second basis $\hat{e}_2$ and $\hat{e}_2^{\prime}$. In higher dimensions, there may exist arbitrariness for the basis, so we decide the second basis by making $\theta_2$ as small as possible and use the principle to determine the rest basis. }
    \label{fig:p_angle}
\end{figure}

\begin{figure}[tbp]
    \centering
    \includegraphics[width=0.45\textwidth]{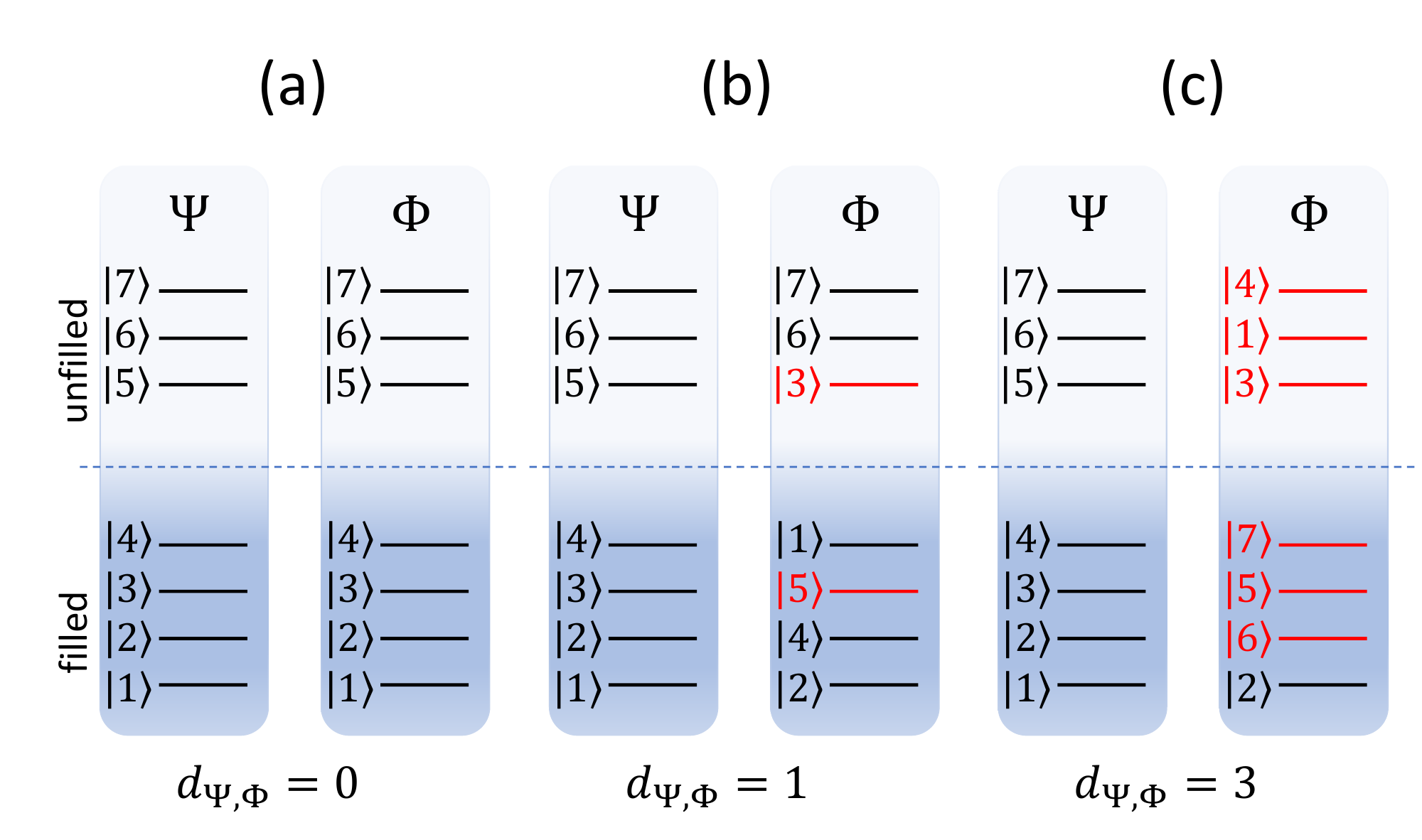}
    \caption{Illustrations of quantum disparity of two systems $\Psi$ and $\Phi$ for $4$ filled levels of seven. $\left\{\ket{l} \right\}_{l=1}^7$ is an orthonormal basis. Red colors highlight the differences between $\Psi$ and $\Phi$. The quantum disparity $d_{\Psi,\Phi}$ is independent of the order of the filled/unfilled levels; it is equal to the number of different kets in filled levels between $\Psi$ and $\Phi$.}
    \label{fig:basis}
\end{figure}

Suppose that there exists another multilevel system $\Phi = \left\{ \ket{\phi_1},\ket{\phi_2},\hdots,\ket{\phi_k} \right\}$ in the same Hilbert space with different $k$ filled states, and this system is considered another $k$-plane. 
The quantum disparity of $\Psi$ and $\Phi$ is related to the distance in the Grassmann manifold, which is closely related to the relative angle of the two $k$-planes. Instead of $k(k-1)/2$ relative angles among $\ket{\psi}$'s and $\ket{\phi}$'s, the informative principal angles are defined by, with a proper choice of coordinates,
\begin{equation}
    \cos \theta_i = \max_{\psi'_i \in \Psi,~\phi'_i \in \Phi}\braket{\psi'_i}{\phi'_i}, \quad i=1,\hdots,k.
\end{equation}
with $\braket{\psi_i'}{\psi_j'} = \braket{\phi_i'}{\phi_j'} = \delta_{ij}$. An example of the principal angles of two planes in three dimensions is seen in Fig. \ref{fig:p_angle}, which illustrates the principal angles. Specifically, we evaluate the inner product of $\Psi$ and $\Phi$, a $k \times k$ matrix,
\begin{equation}
    \Psi^{\dagger} \Phi = 
    \begin{pmatrix}
        \braket{\psi_1}{\phi_1} & \braket{\psi_1}{\phi_2} & \hdots & \braket{\psi_1}{\phi_k} \\
        \braket{\psi_2}{\phi_1} & \braket{\psi_2}{\phi_2} & \hdots & \braket{\psi_2}{\phi_k} \\
        \vdots & & \vdots \\
        \braket{\psi_k}{\phi_1} & \braket{\psi_k}{\phi_2} & \hdots & \braket{\psi_k}{\phi_k} \\
    \end{pmatrix}
    = V \Sigma U^{\dagger},
\end{equation}
and perform its singular value decomposition ($\Sigma$ a semipositive diagonal matrix and $U,~V \in U(k)$). The singular values $s_i = \Sigma_{ii} =\cos \theta_i $ are interpreted as the cosine of the principal angles for $0 \le s_i \le 1$. The functions of $U$ and $V$ are to rotate the coordinate systems in two $k$-planes, such that the resulting base sets have the greatest parallelism. Similar to Eq.~(\ref{duv}), the quantum disparity of the multiband systems is defined as
\begin{equation}
    d_{\Psi,\Phi} =  \frac{2}{\pi} \sum_{i=1}^{k} \arccos{s_i}=  \frac{2}{\pi} \sum_{i=1}^{k} \theta_{i}. 
\end{equation}
Geometrically, if two $k$-planes share a common $l$-subplane, $l\le k$, where $\theta_i =0$, the quantum disparity is determined by the remaining ($k-l$)-subplanes. Schematic illustrations for the quantum disparity of integral values are shown in Fig.~\ref{fig:basis}, where the quantum disparity is determined by the number of bases in which the two systems differ. Therefore, in our definition, the quantum disparity between two states in $G_{k,n}(\mathbb{C})$ is in the range: 
\begin{equation}
    0 \le d \le \min(k,n-k).
\end{equation}

We provide another definition of the Grassmann manifold in terms of projectors. The projector onto $\Psi$ is $P_{\Psi} = \Psi \Psi^{\dagger} = \sum_{i=1}^{k} \ketbra{\psi_i}$. (The projector is a Gram matrix, whose entries for a reference basis $\left\{\ket{\alpha} \right\}_{\alpha = 1}^n$ are given by the inner product: $\left( P_{\Psi} \right)_{\alpha \beta} = \expval{\vb{v}_\alpha,\vb{v}_\beta}$, where $(\vb{v}_\beta)_i := \braket{\psi_i}{\beta} $.) The advantage of using projectors for Grassmannian elements is that projectors are invariant under any unitary transformation among filled states (and unfilled states separately). Therefore, two states are unitarily equivalent if and only if their projectors are equal. The projectors have these properties: (i) they are Hermitian (e.g. $P_{\Psi}^{\dagger} = P_{\Psi}$), (ii) their squares return to themselves ($P_{\Psi}^2 = P_{\Psi}$), and (iii) their ranks are both $k$ ($\rank P_{\Psi} = \rank P_{\Phi} =  k$). The Grassmannian $G_{k,n}(\mathbb{C})$ is defined as all collections of matrices of these properties. The Euclidean inner product of two projectors is $ \tr \left(P_{\Psi}^{\dagger} P_{\Phi}\right) = \sum_{ij} \abs{\braket{\psi_i}{\phi_j}}^2 = \tr \left( V \Sigma^2 V^\dagger\right) = \sum_{i=1}^{k} s_i^2$. As a result, the quantum disparity of two projectors is defined as 
\begin{equation}
    d_{\Psi,\Phi} = \frac{2}{\pi} \tr \left( \arccos \sqrt{ \abs{P_{\Psi} P_{\Phi}}} \right),
\end{equation}
where $\abs{P_{\Psi} P_{\Phi}} = \sqrt{P_{\Psi} P_{\Phi} \left(P_{\Psi} P_{\Phi} \right)^\dagger}$.

We note that there are many ways to measure the distance between two matrices. A commonly adopted quantum distance is the Hilbert–Schmidt distance (Frobenius norm of the difference of two matrices) 
\begin{equation}
    D_{\text{HS}}(\Psi,\Phi) = \frac{1}{\sqrt{2}}\| P_{\Psi}-P_{\Phi} \|_{\text{F}} =\sqrt{\frac{1}{2}\tr \left(P_{\Psi} - P_{\Phi}\right)^2} ,
\end{equation}
which gives $D_{\text{HS}}(\Psi,\Phi)=\sqrt{k - \sum_{i} s_i^2 } = \sqrt{ \sum_{i=1}^{k} \sin^2 \theta_i }$. For a single-level system, it is $D_{\text{HS}} = \sqrt{1 - \abs{\braket{\psi}{\phi}}^2} = \sin \theta_{\psi,\phi}$. 
Both the disparity and the distance can quantify the similarity of two systems, but only $D_{\text{HS}}$ is a sensible distance in Euclidean geometry (see Appendix \ref{App_A}) and the quantum disparity does not hold the triangle inequality property. However, $d_{\Psi,\Phi}$ can provide an immediate picture of the difference between two quantum systems. 
When the quantum distances between single-particle states across momentum space are analyzed, a clearer understanding of the topology is achieved. 

In particular, the Hilbert-Schmidt distance of two states can be expressed as the distance of two geometric vectors in Euclidean space: $D_{\text{HS}}(\Psi,\Phi) = \| \vb{V}_{\Psi} - \vb{V}_{\Phi} \|$ (shown in Appendix \ref{App_A}), so we can take the vector as the position of the quantum state and the collection of positions forms the manifold of the system. This is a way of embedding an abstract manifold in Euclidean space. We will construct the manifold from the table of quantum distances by using the classical multidimensional scaling (CMS) technique. (The details of CMS are provided in Appendix \ref{App_B}.) CMS is somewhat similar to principal component analysis; it can determine the rank of the vector space and consequently answer the minimal dimension of the embedding space.

\section{Results} \label{sec:results}
\subsection{2D Chern insulators} \label{sec:2D_Chern}
\subsubsection{Two-band systems}
In two spatial dimensions, an insulator without time-reversal symmetry (in symmetry class A) is classified by an integer, named the Chern number, due to the homotopy group $\pi_2[G_{k,n}(\mathbb{C})]=\mathbb{Z}$. The Chern number $C$ is the integral of the Berry curvature in the first Brillouin zone (BZ): 
\begin{equation}
    C = \frac{1}{2\pi}\int \dd[2]{k} \left[\partial_{x} A_y(\bk) - \partial_{y} A_x(\bk) \right],
\end{equation}
where $A_{j}(\bk)=i \sum_{n\in \text{filled}} \expval{\partial_{j}}{u_n(\bk)} $, with $\partial_{j} = \partial/\partial k_{j}$ for short, is the Berry connection of the valence bands. This type of topological insulators is called a Chern insulator. The simplest two-band Hamiltonian is expressed as 
\begin{equation}
H^{(n)}(\bk) = \text{Re} \left[ h_{+}^{(n)}(\bk) \right]  \sigma_1 + \text{Im} \left[ h_{+}^{(n)}(\bk)\right] \sigma_2 + h_z(\bk) \sigma_3,
\label{H2d}
\end{equation}
where 
\begin{align}
\begin{split}
    h_{+}^{(n)}(\bk) &= \left(\sin k_x + i \sin k_y \right)^n 
    \quad (n \in \mathbb{N}),  \\
    \quad h_z(\bk) &=m_0 - \cos k_x - \cos k_y.
\end{split}
\end{align}
The lattice constants are set to unity throughout the paper. The momentum is confined to the BZ, $k_x,~k_y \in (-\pi,\pi]$, forming a torus $T^2$. The Hamiltonian models an insulator with a Chern number $C=n$ when $0<m_0<2$. To achieve a negative Chern number, one can replace $h_{+}^{(n)}$ with its complex conjugate in the Hamiltonian. Writing $H^{(n)}=|H^{(n)}| \hat{n} \cdot \vec{\sigma}$, with $|H^{(n)}| = \sqrt{|h_{-}^{(n)}|^2 + h_z^2}$, the system is interpreted as a spin in a $k$-dependent magnetic field along $\hat{n}(\bk)$. We identify the spin for the valence (conduction) band as $\hat{n}(\bk)$ since it is antiparallel (parallel) to the magnetic field due to $\hat{n}(\bk) \ket{u(\bk)} = -(+) \ket{u(\bk)}$. The image $\hat{n}(\bk)$ lies on a sphere $S^2$, consistent with isomorphisms $G_{1,2}(\mathbb{C}) \cong \mathbb{C} P^1 \cong S^2$. The number of times $\hat{n}(\bk)$ wraps the BZ $T^2$ with a sense of direction along $\bk$ gives the Chern number. Note that the quantum distance or disparity cannot tell the orientation of the manifold.

The quantum distance can reveal the geometry and topology of the image manifold. Consider a reference state with some momentum as a point on $S^2$, say the north pole. If there exists a state at a different momentum that has zero distance (or disparity) from the reference state, this state will also be located at the north pole. Conversely, if another state has the maximum disparity of $d=1$ from the reference state, it is located at the south pole. When this occurs, it is expected that one can find a set of states forming a ring on the equator for $d=0.5$ since the manifold is continuous. From a topological perspective, the necessary condition for a non-trivial wrapping over $T^2$ is the existence of a state in the BZ with $d=1$. We will show that the absolute value of the Chern number is determined by the number of $d=1$ states in the BZ.

\begin{figure}[tb]
    \centering
    \includegraphics[width=0.45\textwidth,trim={3cm 2cm 3cm 1cm},clip]{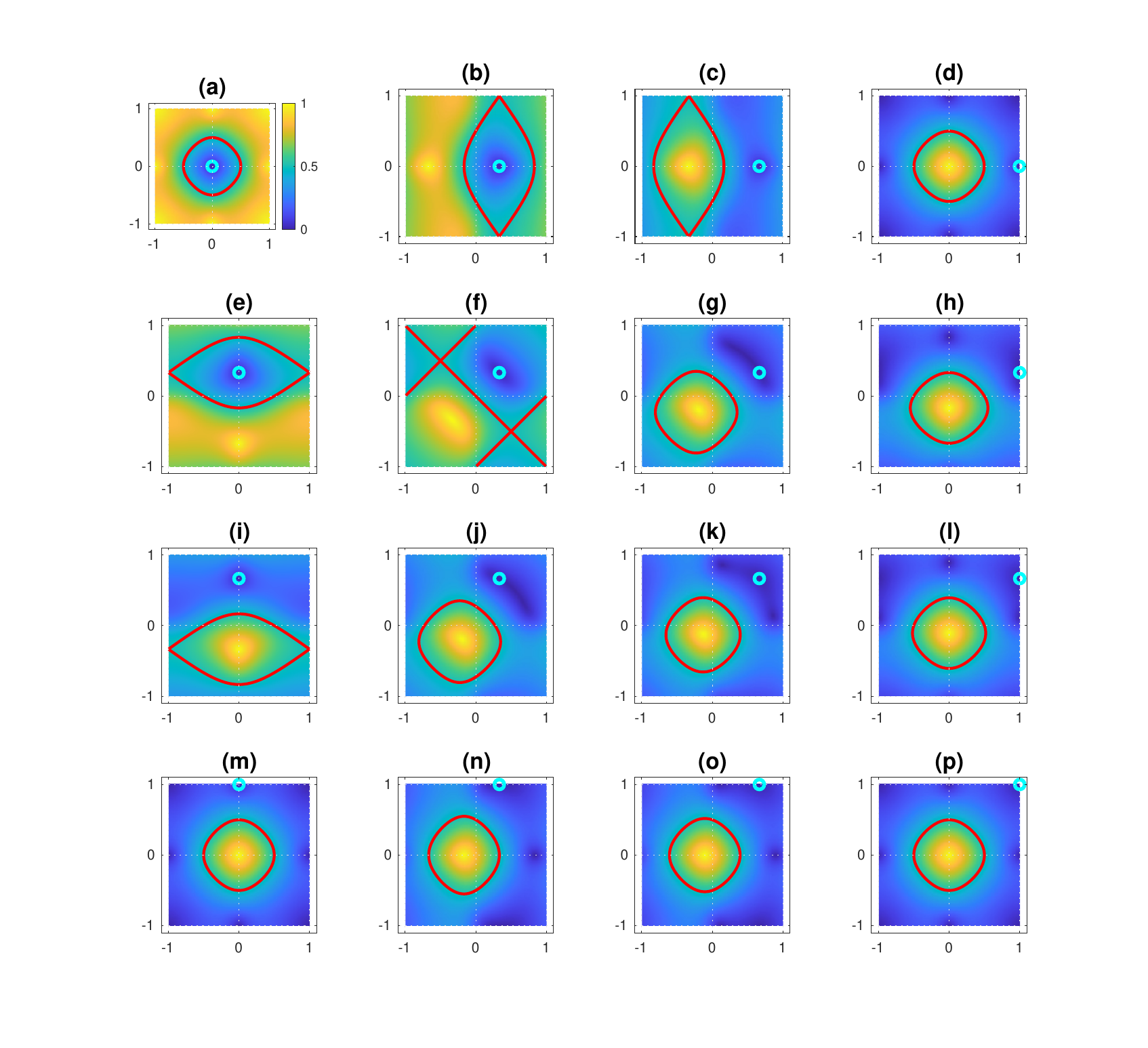}
    \caption{Maps of quantum disparity $d_{\bk,\bk'}$ for the two-band Chern insulator with $C=1$ in Eq. (\ref{H2d}). In each subfigure, the range of $\bk$ is the first BZ and the cyan circle marks the reference point $\bk'$. The red lines highlight the contours of $d=1/2$. The $x$ and $y$ axes are for coordinates $k_x/\pi$ and $k_y/\pi$, respectively.}
    \label{fig:c1}
\end{figure}

\begin{figure}[tb]
    \centering
    \includegraphics[width=0.45\textwidth,trim={3cm 2cm 3cm 1cm},clip]{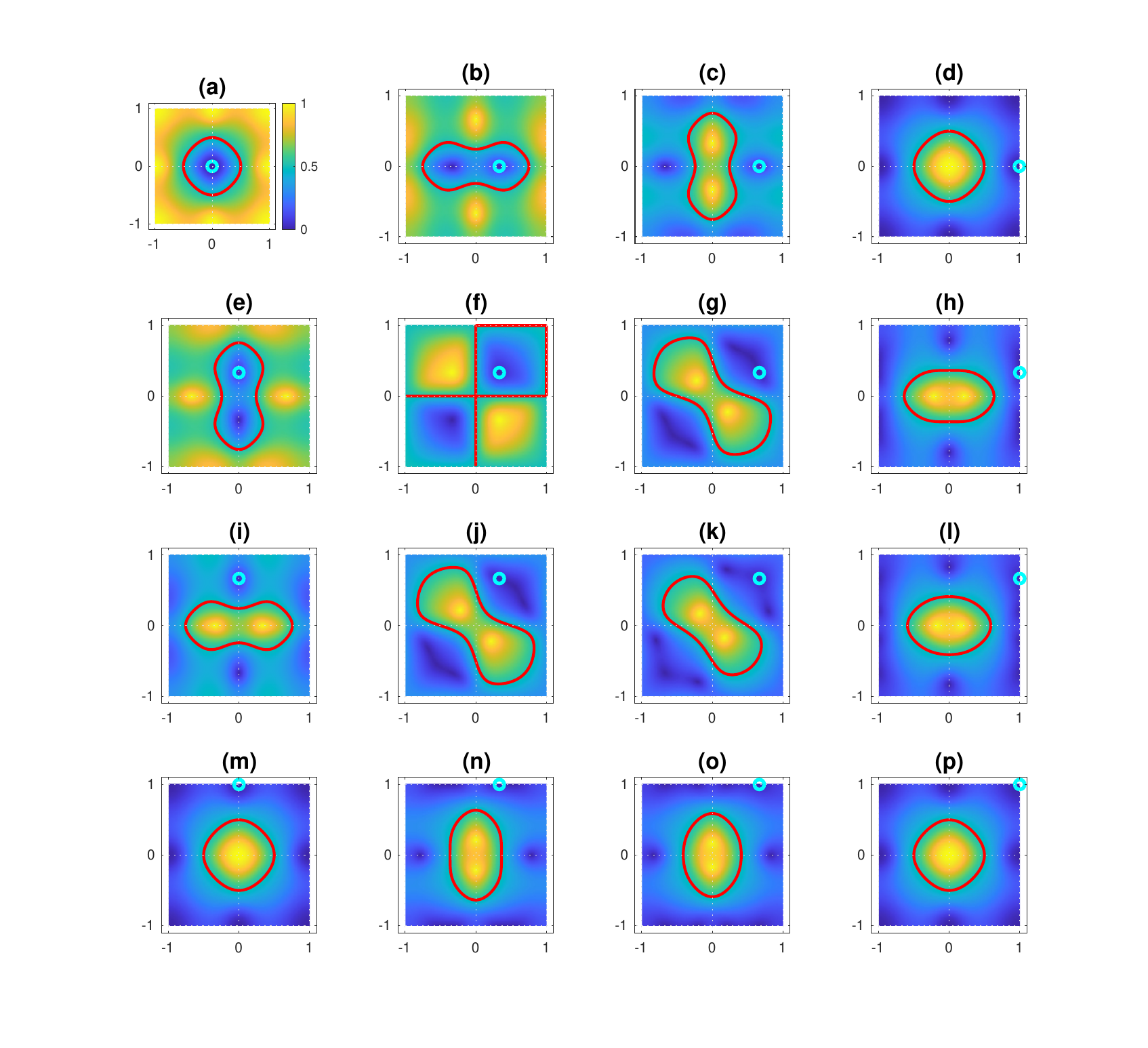}
    \caption{Quantum disparity for $C=2$. Same setting as in Fig. \ref{fig:c1}.}
    \label{fig:c2}
\end{figure}

\begin{figure}[tb]
    \centering
    \includegraphics[width=0.45\textwidth,trim={3cm 2cm 3cm 1cm},clip]{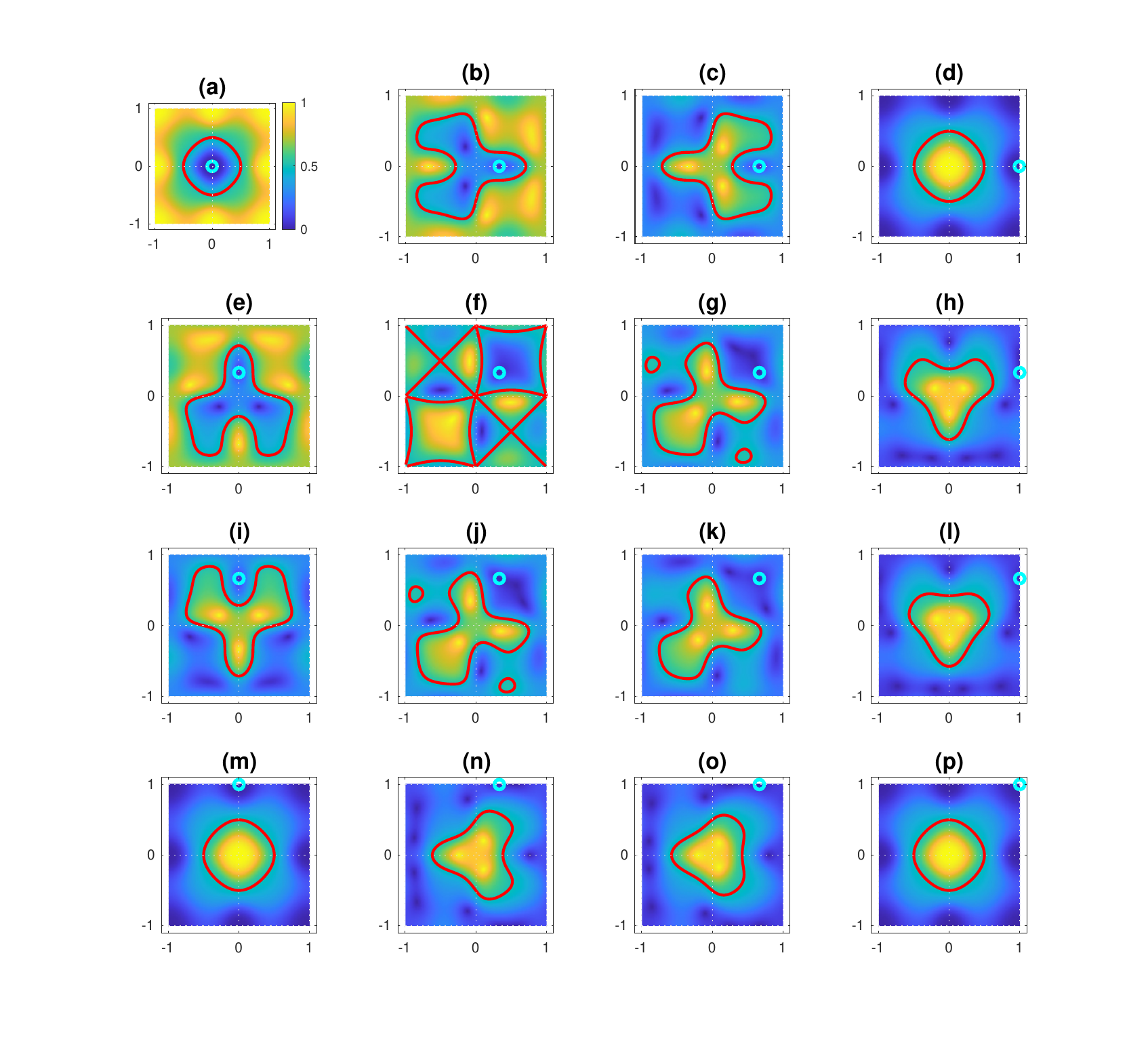}
    \caption{Quantum disparity for $C=3$. Same setting as in Fig. \ref{fig:c1}.}
    \label{fig:c3}
\end{figure}

We demonstrate the quantum disparity $d_{\bk,\bk'}$ of the valence states of Eq.~(\ref{H2d}) with $m_0=1$ at different momenta for $n=1$ in Fig. \ref{fig:c1}, $n=2$ in Fig. \ref{fig:c2}, and $n=3$ in Fig. \ref{fig:c3}. In each subfigure, the reference point $\bk'$ is chosen by the cyan circle and $\bk$ runs over the whole BZ. In this model Eq.~(\ref{H2d}), the states at $\bk=(\pi,0), (0,\pi),\mathrm{and}~(\pi,\pi)$ are identical regardless of the Chern number.

\begin{figure}[tb]
    \centering
    \includegraphics[width=0.45\textwidth]{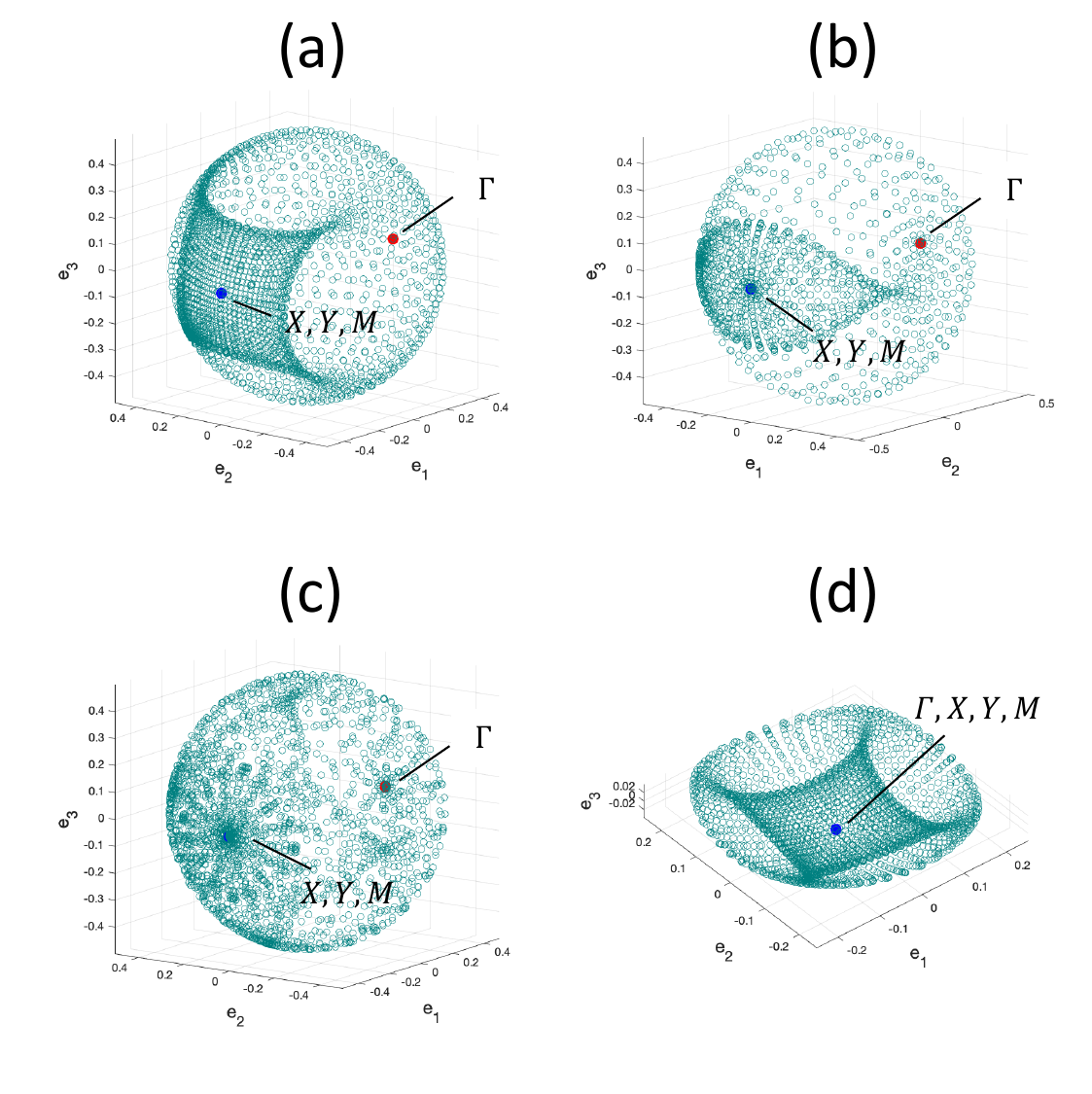}
    \caption{Euclidean manifolds for the two-band Chern insulators based on the Hilbert-Schmidt distance matrices for $C=1$ (a), $C=2$ (b), $C=3$ (c), and $C=0$ when $n=1$ and $m_0=3$ in Eq. (\ref{H2d}) (d). Same setting as in Fig. \ref{fig:c1}.}
    \label{fig:manifold_S2}
\end{figure}

The reconstructed manifolds from $D_{\text{HS}}(\bk,\bk')$ by CMS of the two-band system with different Chern numbers are shown in Fig. \ref{fig:manifold_S2}. It is evident that the manifolds for topological insulators are $S^2$ in Figs. \ref{fig:manifold_S2}(a), (b), and (c). An ideal round $S^2$ indicates that any state in $\bk$ space has its antipodal partner with $d=1$. As for the $C=0$ trivial insulator, the manifold is a $D^2$ that does not allow a $d=1$ pair. 

\subsubsection{Four-band systems}
We adopt the four-band model from Ref. \cite{alexandradinata2014}
\begin{align}
\begin{split}
     H_{\mathrm{4band}}(\bk)&=\frac{1}{2}\Gamma_{13} - \Gamma_{03} -\frac{3}{4}(\Gamma_{30}+\Gamma_{03})(\cos k_x+ \cos k_y) \\
    &+(\Gamma_{12}+\Gamma_{31})\sin k_x + (\Gamma_{21}+\Gamma_{32}) \sin k_y ,
    \label{H4band}
\end{split}
\end{align}
where Gamma matrices are tensor products of Pauli or identity matrices, $\Gamma_{ij} = \sigma_i \otimes \tau_j$.
The model lacks time-reversal symmetry but possesses inversion symmetry: $\Gamma_{03}H_{\mathrm{4band}}(\bk)\Gamma_{03} = H_{\mathrm{4band}}(-\bk)$. The resulting four bands are separated in energy, with Chern numbers of $-1,+2,-2,\mathrm{and}+1$ from high to low energy, respectively. The lower two are referred to as valence bands, and their combined Chern number is the sum of individual Chern numbers, giving $C=-1$.   

\begin{figure}[tb]
    \centering
    \includegraphics[width=0.45\textwidth,trim={3cm 2cm 3cm 1cm},clip]{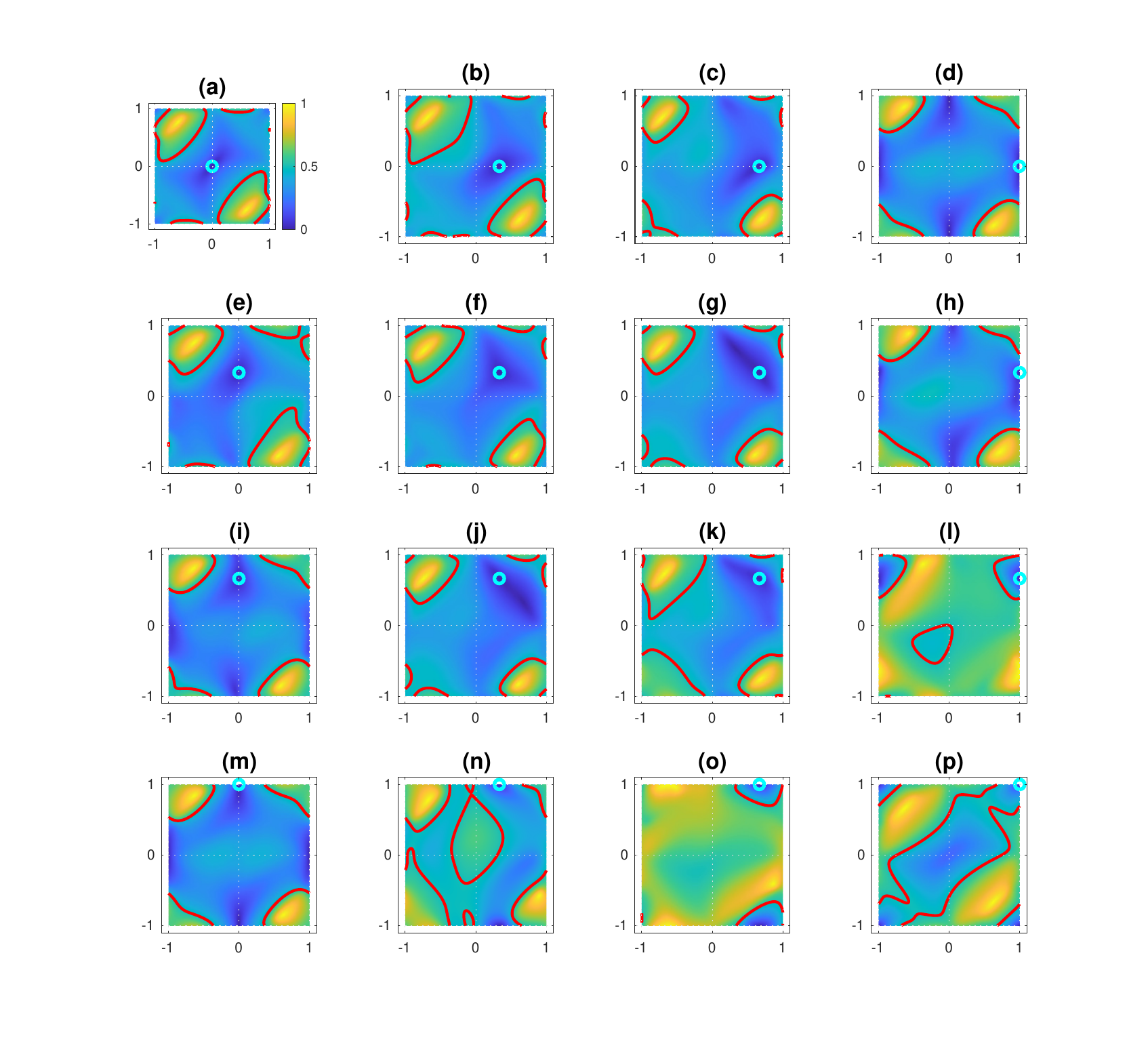}
    \caption{Maps of quantum disparity $d_{\bk,\bk'}$ of states in the top valence band of the four-band model in Eq. (\ref{H4band}). The Chern number from the band is $C=-2$.}
    \label{fig:c2_4band}
\end{figure}

Firstly, we show the quantum disparity in the manifold from the top valence band [$\in G_{1,4}(\mathbb{C})\cong \mathbb{C} P^3$] in Fig. \ref{fig:c2_4band}. It is clear and consistent that the number of $d=1$ $\bk$-points in each subfigure is two, which matches the absolute value of the Chern number. Unlike the two-band model, the figures suggest a complex geometry in the manifold. In the two-band case, two distinct $d=1$ $\bk$-points correspond to an identical state because the dimension of the Hilbert space is only two. In the four-band case, it is possible for two $d=1$ $\bk$-points to represent different states due to higher degrees of freedom. This was verified; for example, the two states at $\pm (-0.59\pi,0.76 \pi)$ [yellow apexes in Fig.~\ref{fig:c2_4band}(a)] that are orthogonal to the $\bk'=0$ state are neither identical nor orthogonal to each other, with a quantum disparity of $d=0.7711$. 

\begin{figure}[tb]
    \centering
    \includegraphics[width=0.45\textwidth,trim={3cm 2cm 3cm 1cm},clip]{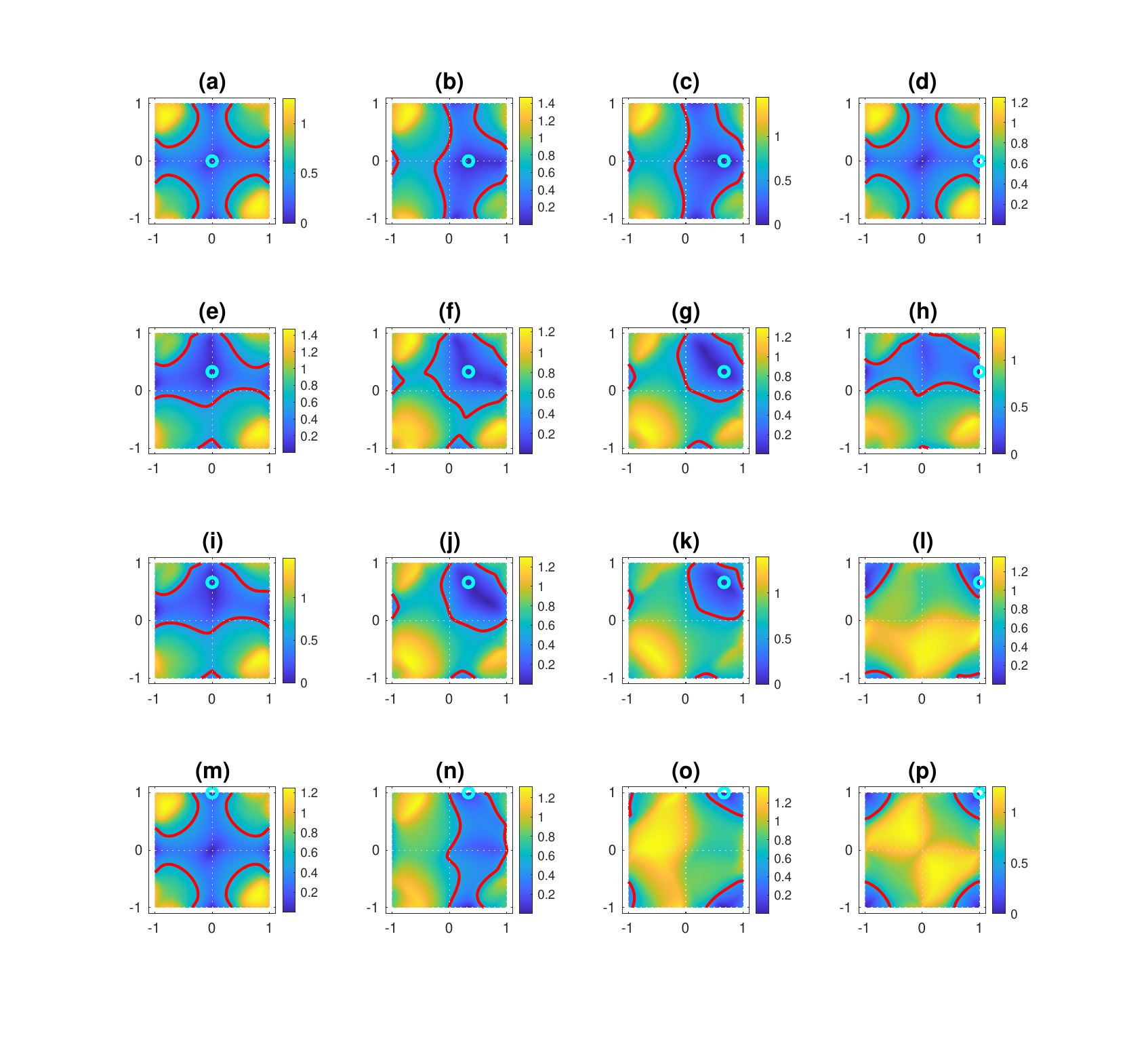}
    \caption{Maps of quantum disparity $d_{\bk,\bk'}$ of states in the two-valence bands of the four-band model in Eq. (\ref{H4band}). The Chern number of the bands is $C=-1$. The cyan circle marks the reference point $\bk^{\prime}$. }
    \label{fig:c1_4band}
\end{figure}

Secondly, we show the quantum disparity in the manifold of the two valence bands [$\in G_{2,4}(\mathbb{C})$] in Fig. \ref{fig:c1_4band}. Unlike the one-band case, the quantum disparity can exceed one (but not not two). The Chern number now seems not to match the number of $d=1$ states. This is because the quantum disparity is the sum of two principal angles (divided by $\pi/2$). A better approach is to count the number of basis changes. If we only consider the maximum of the two principal angles, the number of the maximal angles equal $\pi/2$ indeed corresponds to the Chern number (not shown).

\begin{figure}[tb]
    \centering
    \includegraphics[width=0.5\textwidth]{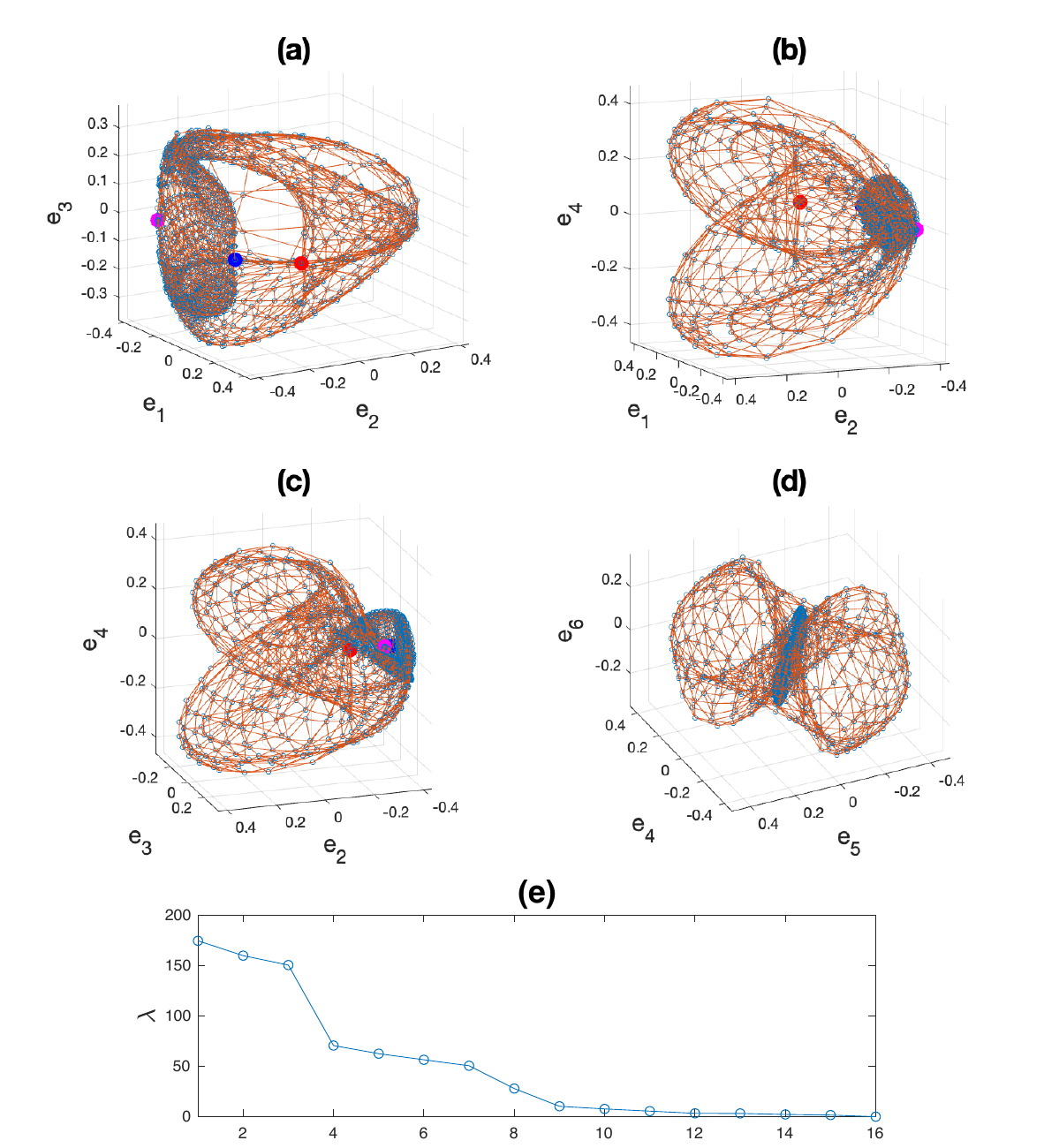}
    \caption{Manifold for the $C=-2$ top valence band of the four-band model, which quantum parity is in Fig.~\ref{fig:c2_4band}. The eigenvalues of the Gram matrix indicate the dimension of the space is 15 (e). We select some coordinate bases and show the manifold projection in these coordinates. Two-dimensional closed surfaces with a void at the center are seen in (b) and (c). The blue, red, and magenta solid circles are for states at $\Gamma$, $M$, and $X (Y)$, respectively. }
    \label{fig:manifold_2D_4band_Ch2}
\end{figure}

\begin{figure}[tb]
    \centering
    \includegraphics[width=0.5\textwidth]{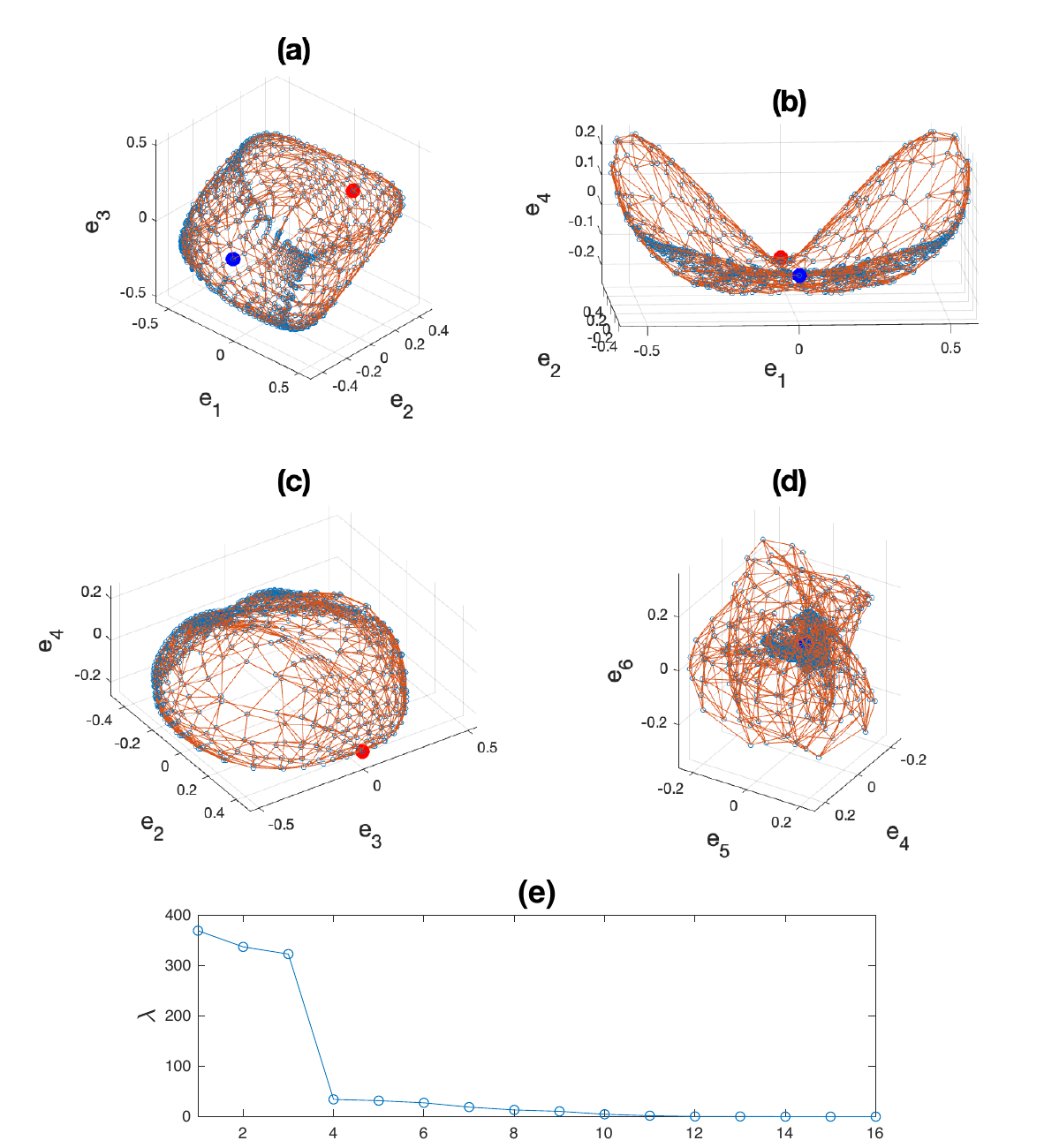}
    \caption{Manifold for the two valence bands of the four-band model, $C=1$, which quantum parity is in Fig.~\ref{fig:c1_4band}. (e) shows that the major components are the first three coordinates. It is evident in (a) to see a closed $S^2$ when projecting the manifold in the first three coordinates.}
    \label{fig:manifold_2D_4band_Ch1}
\end{figure}

The manifolds for the two systems are shown in Figs.~\ref{fig:manifold_2D_4band_Ch2} and \ref{fig:manifold_2D_4band_Ch1}, respectively. As the maps of quantum disparity show, the geometry of the manifolds is complex. Both manifolds are embedded in 15 dimensions as seen in Panel (e). Visualizing a high-dimensional object is challenging. For each plot, we show the projection of positions in a three-dimensional coordinate frame; for example, projecting a point on the $e_1$-$e_2$-$e_3$ frame transforms $(x_1,x_2, x_3, ...,x_{15}) \to (x_1,x_2, x_3)$. We believe that since the topology is classified by the second Homology group $H_2$ the method can portray the topology.

For the $C=-2$ system, we see closed surfaces enclosing the origin in Fig. \ref{fig:manifold_2D_4band_Ch2}(b) and (c). These closed surfaces have voids (``holes") in the center, which are topologically equivalent to $S^2$. The possibility of an $S^3$ is excluded because the projection of an $S^3$ in 3D is a 3-ball (filled sphere). Notably, we also see two wraps for $C=-2$ for two misaligned $S^2$'s.  
For the $C=1$ system, in Fig. \ref{fig:manifold_2D_4band_Ch2}, the manifold appears simpler because the relevant dimensions are the first three, as shown in the eigenvalue spectrum in Fig. \ref{fig:manifold_2D_4band_Ch2}(e). In these three dimensions, an $S^2$ for $C=1$ is evident in Fig. \ref{fig:manifold_2D_4band_Ch2}(a).

\subsection{2D topological insulators} \label{sec:2D_TI}
The 2D topological insulator in class AII is modeled as
\begin{align}
    \begin{split}
        H_{\mathrm{2DTI}}(\bk) &= \tau_1 \sigma_1 \sin k_x + \tau_1 \sigma_2 \sin k_y \\
        &+ \tau_3 \sigma_0 (m_0 -\cos k_x-\cos k_y) + \Delta \tau_1 \sigma_0,
    \end{split}
\end{align}
where $0<|m_0|<2$ and $\Delta$ is to break inversion symmetry. The system is protected by time-reversal symmetry and is classified by a $\mathbb{Z}_2$ topological invariant ($\nu=1$). When $\Delta=0$, each band has two-fold degeneracy imposed by the coexistence of time-reversal and inversion symmetries, which operators are $\mathcal{T} = i \tau_0 \sigma_y$ and $\mathcal{I}=\tau_3 \sigma_0$, respectively. When $\Delta$ is finite, inversion symmetry is broken and the spin degeneracy is lifted in bands except at time-reversal invariant momenta. 

For $\Delta=0$, the quantum disparity profile is identical to the $C=1$ Chern insulator in Fig. \ref{fig:c1} but doubles the value. The doubling is due to degeneracy two both in conduction and valence bands. Therefore, the maximal quantum disparity is two, implying two coincident basis exchanges between the valence and conduction bands. 

\begin{figure}[tb]
    \centering
    \includegraphics[width=0.45\textwidth,trim={3cm 2cm 3cm 1cm},clip]{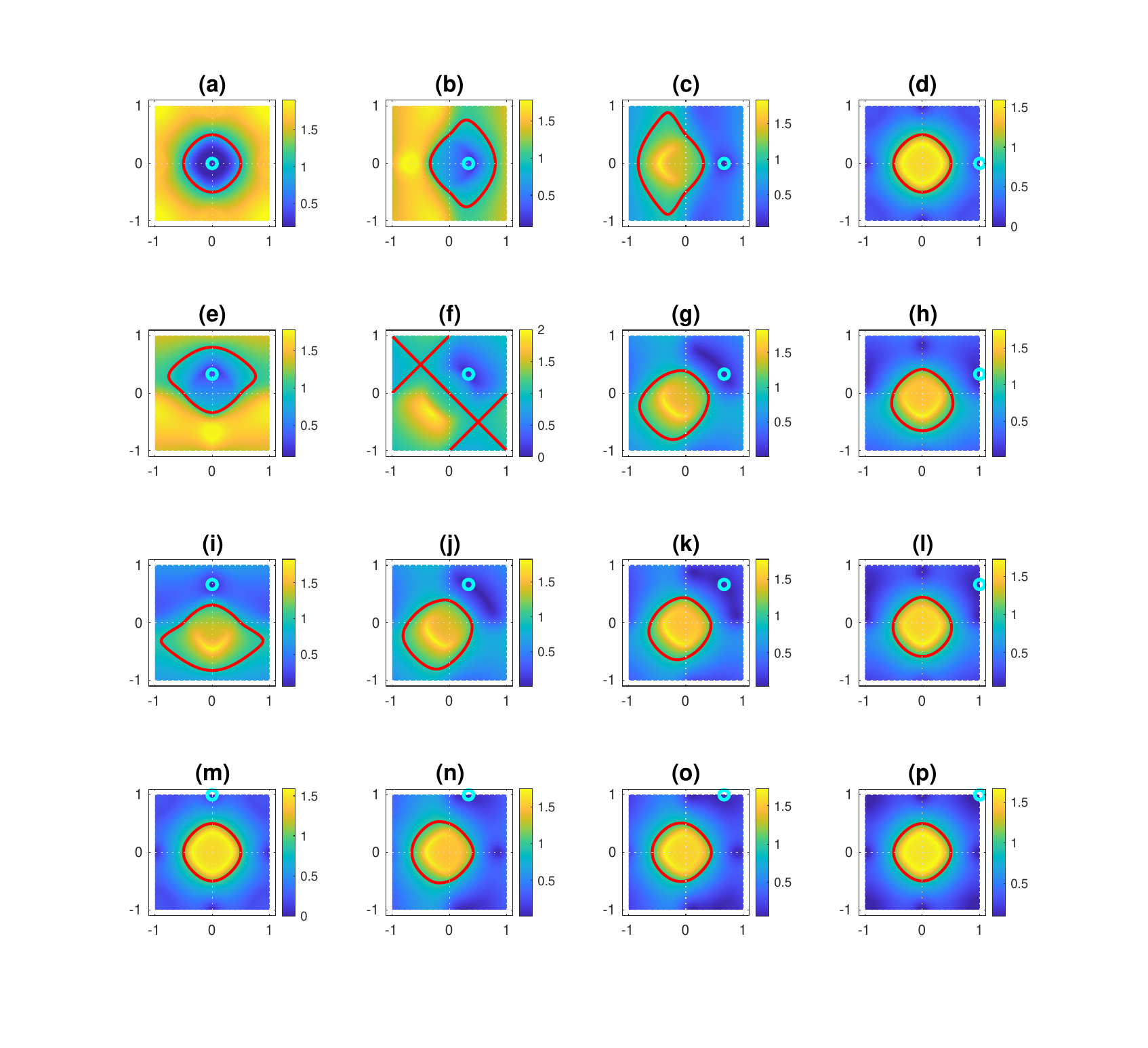}
    \caption{Maps of quantum disparity for a 2DTI with $m_0=2$ and $\Delta = 0.6$. Red lines for $d=1$ contours.}
    \label{fig:2DTI}
\end{figure}

\begin{figure}[tb]
    \centering
    \includegraphics[width=0.45\textwidth,trim={3cm 2cm 3cm 1cm},clip]{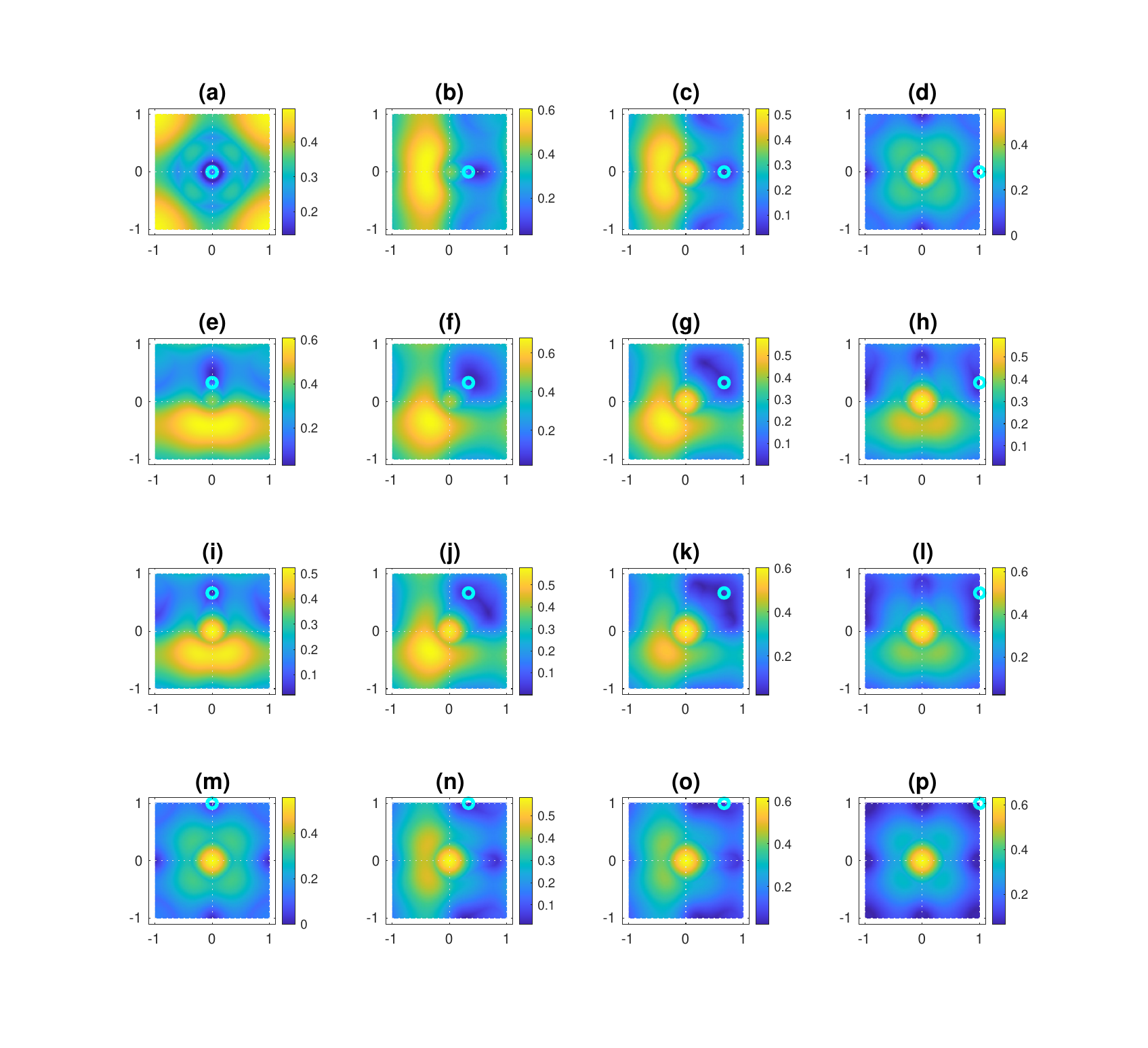}
    \caption{Maps of quantum disparity for the trivial 2D time-reversal insulator with $m_0=3$ and $\Delta = 0.6$. Red lines for $d=1$ contours.}
    \label{fig:2DTI_0}
\end{figure}

It would be more important to understand the system without inversion symmetry, that is $\Delta \neq 0$, since two band exchanges happen at different momenta. We show the result in Fig. \ref{fig:2DTI} with $m_0=1$ and $\Delta = 0.6$. It shows that the quantum disparity cannot reach $d=2$ as the $\Delta = 0$ case. However, the maximal $d$ is larger than one and its location is not at discrete $\bk$ points but somewhat over an area. The existence of $d>1$ at all $\bk$ indicates the non-triviality. To confirm that, we compare the quantum disparity in a trivial insulator ($m_0=3$ and $\Delta = 0.6$) in Figure \ref{fig:2DTI_0}. It is clear that for a time-reversal insulator, the quantum disparity cannot be greater than $1$ for all $\bk$. 

\subsection{Hopf insulator} \label{sec:hopf}
In three dimensions, class A allows a special homotopy for two-band systems, giving a $\mathbb{Z}$-classification by the Hopf invariant 
\begin{equation}
    N_h=- \frac{1}{2\pi}\int \dd[3]{k} \vb{A}(\bk)\vdot \vb{J}(\bk),
\end{equation}
where the topological current density is $\vb{J} = \grad_{\bk} \cross \vb{A}$. The Hopf map, $S^3 \rightarrow S^2$, implies that $S^3$ is a $U(1)$ bundle over $S^2$. Thus, the preimage of every point on $S^2$ will be a ring in $S^3$. The Hopf invariant is the linking number of preimages of antipodal points of the $S^2$ \cite{wilczek1983,moore2008,kennedy2016}. A Hamiltonian for a $N_h=1$ Hopf insulator is described by $H(\bk) = d_x(\bk) \sigma_1 + d_y(\bk) \sigma_2 + d_z(\bk) \sigma_3 $, where
\begin{align} \label{dxyz}
\begin{split}
d_x(\bk) &= 2 \sin (k_x) \sin (k_z) + 2 \sin (k_y) M(\bk), \\
d_y(\bk) &= -2 \sin (k_y) \sin (k_z) + 2 \sin (k_x) M(\bk), \\
d_z(\bk) &=  \sin[2](k_x) + \sin[2](k_y) - \sin[2](k_z)- M^2(\bk).
\end{split}
\end{align}
with $M(\bk)=\sum_{i=x,y,z} \cos (k_i) - m_0$. To be topological, the value of $m_0$ has to be $1<\abs{m_0}<3$~\cite{moore2008}.

\begin{figure}[tb]
    \centering
    \includegraphics[width=0.45\textwidth]{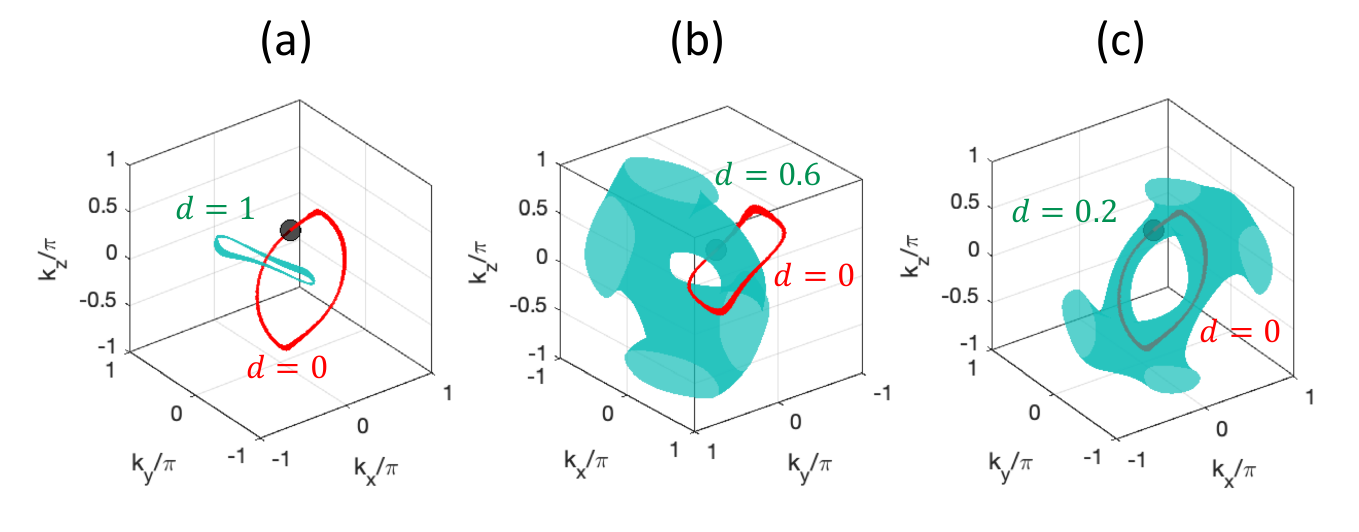}
    \caption{Hopf links in the Hopf insulator with $m_0=2$. Green regions are points $\bk$ with (a) $d_{\bk,\bk^{\prime}}=1$, (b) $d_{\bk,\bk^{\prime}}=0.6$, (c) $d_{\bk,\bk^{\prime}}=0.2$. The red line are $\bk$ with $d_{\bk,\bk^{\prime}}=0$. The black circle is the position of the reference point $\bk'$.}
    \label{fig:Hopflink}
\end{figure}

The manifold of the topological phase is an $S^2$ as that of the two-band Chern insulator. To demonstrate the linking number, we arbitrarily pick a reference $\bk'$ point in the BZ as labeled by the black circles in Fig. \ref{fig:Hopflink} and display the $\bk$ points with $d_{\bk,\bk^{\prime}}=0$, which set is a ring as the red lines in Fig. \ref{fig:Hopflink}. We then show the contours that give $d_{\bk,\bk^{\prime}}=1,~d_{\bk,\bk^{\prime}}=0.6$, and $d_{\bk,\bk^{\prime}}=0.2$ in Fig. \ref{fig:Hopflink}(a), (b), and (c), respectively. As expected, the $d=1$ contour forms another ring (by green) that links with the $d=0$ ring. These two rings visualize the preimages of a point and its antipodal point on $S^2$. When $d$ is lowered to 0.6 or 0.2, the equal-$d$ contour becomes a torus $T^2=S^1 \times S^1$, with one $S^1$ representing the latitudinal circle and the other $S^1$ representing the preimage ring. Nevertheless, the $d=0$ ring is linked with an $S^1$ in the $T^2$. We note that when the system is in the trivial phase by tuning $m_0$, the image $H(\bk)/\abs{H(\bk)}$ cannot cover the entire $S^2$, so the maximal $d$ does not reach $1$ and the linking preimage does not appear. 

\subsection{3D axion insulator} \label{sec:axion}
An axion insulator is a 3D insulator that exhibits a quantized topological magnetoelectric (TME) effect, described by the action \cite{wilczek1987,qi2008,turner2012}
\begin{equation}
    S_{\theta} = \frac{\theta}{2\pi} \frac{e^2}{h} \int \dd{t} \dd[3]{x} \vb{E} \vdot \vb{B}.
\end{equation}
To achieve a quantized effect, $\theta=\pi \mod 2\pi$, candidates can be found in systems that preserve either time-reversal ($\mathcal{T}$) symmetry or inversion ($\mathcal{I}$) symmetry; the candidates in the former class are 3D topological insulators. In an $\mathcal{I}$-symmetric axion insulator, the surface state is gapped if no other symmetry protection is present, resulting in a half-quantum Hall effect. The electric transport on the gapped surface relies on emergent gapless hinge states in lower dimensions. Consequently, axion insulators extend to systems of other crystalline symmetries beyond $\mathcal{T}$ or $\mathcal{I}$ symmetry and are identified as higher-order topological insulators (HOTIs) that host chiral hinge states.

We study the $\mathcal{C}_{4z} \mathcal{T}$-symmetric HOTI model from Refs. \cite{schindler2018,li2020}:
\begin{align}
    \begin{split}
        H_{\mathrm{HOTI}}(\bk) &= M(\bk) \tau_3 \sigma_0 + \Delta_1 \sum_{i=x,y,z} \sin k_i \tau_1 \sigma_i \\
        & + \Delta_2 (\cos k_x- \cos k_y)\tau_2 \sigma_0 + \Delta_3 \tau_1 \sigma_0,
    \end{split}
\end{align}
where $M(\bk)=\sum_{i=x,y,z} \cos (k_i) - m_0$ as defined before, and $\Delta_1 = \Delta_2 = 1$ are chosen since their nonzero values do not affect the topology. When $\Delta_2 =\Delta_3=0$, the model becomes a 3D $\mathcal{T}$-symmetric TI if $1 < \abs{m_0} < 3$. A nonzero $\Delta_2$ will break $\mathcal{T}=i \tau_0 \sigma_2 K$, $\mathcal{I}=\tau_3 \sigma_0$, and the fourfold rotation symmetry $\mathcal{C}_{4z}= \tau_0 e^{-i(\pi/4) \sigma_3}$ but retains the combined symmetries $\mathcal{IT}$ and $\mathcal{C}_{4z} \mathcal{T}$. $\mathcal{IT}$ symmetry ensures that every band is doubly degenerate, but this degeneracy is broken by the $\Delta_3$ term, except at eight time-reversal invariant momenta, $k_{i \in \lbrace x,y,z \rbrace }=0$ or $\pi$. 
The anti-unitary nature of $\mathcal{C}_{4z} \mathcal{T}$ is crucial. Since $\left(\mathcal{C}_{4z} \mathcal{T} \right)^4=-1$ for the half-integer spin, all bands show twofold degeneracy at $\mathcal{C}_{4z} \mathcal{T}$-invariant momenta, labeled $\Gamma,M,Z,A$. The wavefunction at these momenta will determine the system's topology.

The topological invariant for the axion insulator is the Chern-Simons form 
\begin{equation}
    \theta = \frac{1}{4\pi} \int \dd[3]{k} \epsilon^{\mu \nu \rho} \tr\left[ A_{\mu} \partial_{\nu}A_{\rho}-i \frac{2}{3}A_{\mu}A_{\nu}A_{\rho} \right],
\end{equation}
which is $\mathcal{C}_{4z} \mathcal{T}$ invariant and thus quantized to $\theta=0,\pi$, gauge invariant up to a multiple of $2\pi$. With some effort \cite{schindler2018}, the Chern-Simons invariant $\theta/\pi$ is equivalent to the winding number of the unitary sewing matrix $B_{mn}(\bk) = \mel{u_{m}(C_{4}T \bk)}{\mathcal{C}_{4z} \mathcal{T}}{u_{n}(\bk)} \in \text{U}(2)$ in the BZ,
\begin{equation}
    2 P_3 = -\int \frac{\dd[3]{k}}{24\pi^2} \epsilon^{\mu \nu \rho} \tr \left[(B \partial_{\mu} B^{\dagger})
    (B \partial_{\nu} B^{\dagger}) (B \partial_{\rho} B^{\dagger})\right],
\end{equation}
where $P_3$ is the magnetoelectric polarization. The U(1) phase in the sewing matrix does not contribute to the winding number, or equivalently we can adiabatically deform the U(2) matrix to an SU(2) one~\cite{wang2010}, so the sewing matrix maps from $T^3$ to $ \text{SU}(2) \cong S^3$. We note that two sewing matrices of opposite signs $\pm B_{mn}(\bk)$ give an identical $P_3$, indicating that the topological invariant is the parity of the wrapping times that $T^3$ covers $\text{SU}(2) /\mathbb{Z}_2 \cong \text{SO}(3)$. The manifold of $\text{SO}(3)$ is a 3-ball $B^3$ with radius $\pi$ and every point on the boundary $S^2$ of the $B^3$ is identified with its antipodal point on $S^2$. Consequently, the manifold is doubly connected: a closed curve connecting a point on the boundary and its antipodal through the origin is a loop that is incontractible, unlike a contractible loop that does not reach the boundary. Additionally, when any curve winds twice, it can continuously shrink to a single point in $B^3$, displaying a $\mathbb{Z}_2$ feature.

\begin{figure}[tb]
    \centering
    \includegraphics[width=0.45\textwidth]{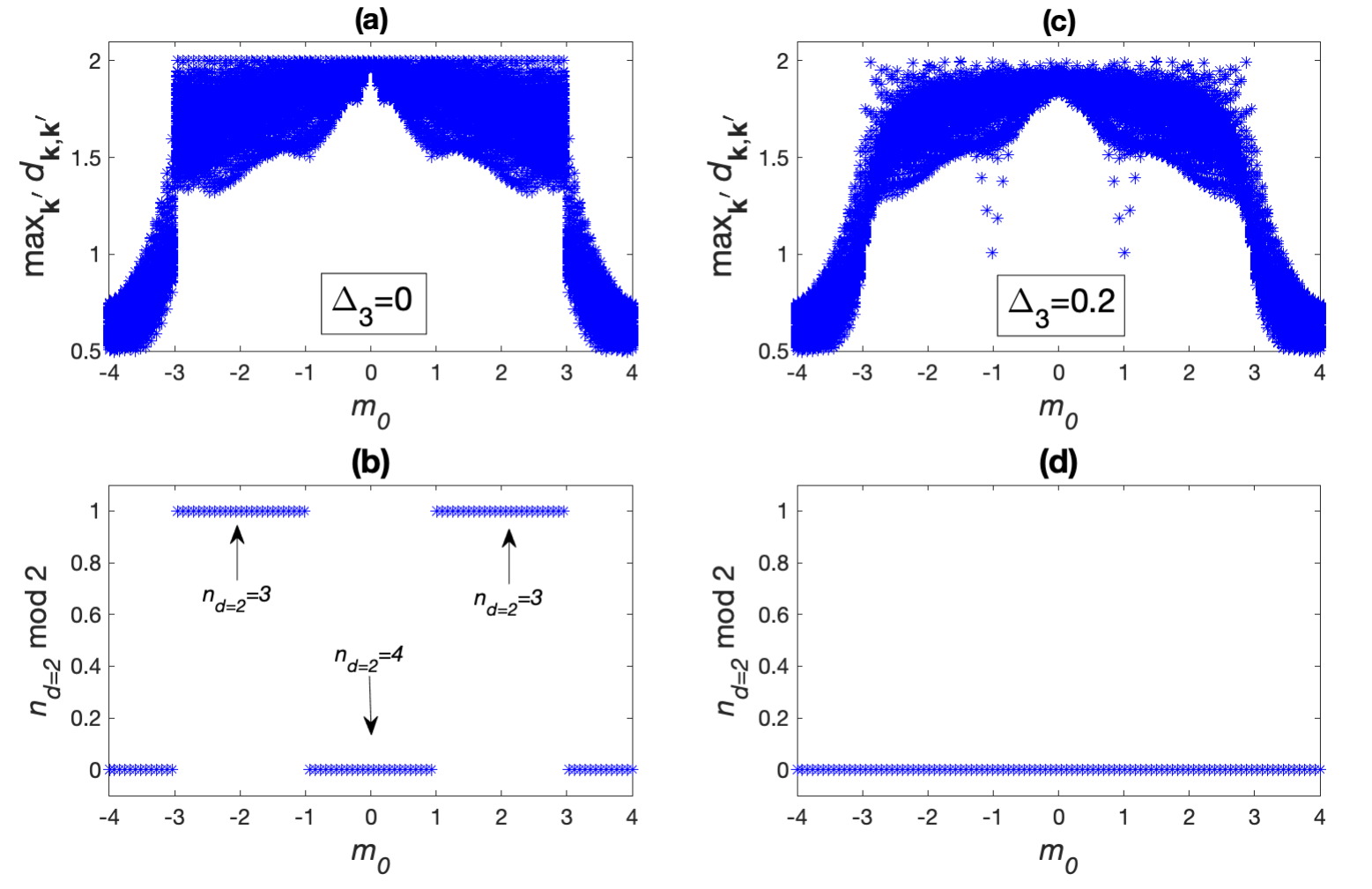}
    \caption{Quantum disparity of the HOTI model with $\Delta_3 =0$ [(a) and (b)] and $\Delta_3 =0.2$ [(c) and (d)]. (a) and (c) show the collection of maximal $d_{\bk,\bk'}$ for all $\bk$ in the BZ for different $m_0$. (b) and (d) show $n_{d=2}$, the half number of $\bk$ points that can find at least a $d=2$ partner $\vb{k}'$. When modulo 2 operation is taken, $n_{d=2}$ is identified with the Chern-Simons invariant $\theta/\pi$ .}
    \label{fig:HOTI}
\end{figure}

\begin{figure}[tb]
    \centering
    \includegraphics[width=0.5\textwidth]{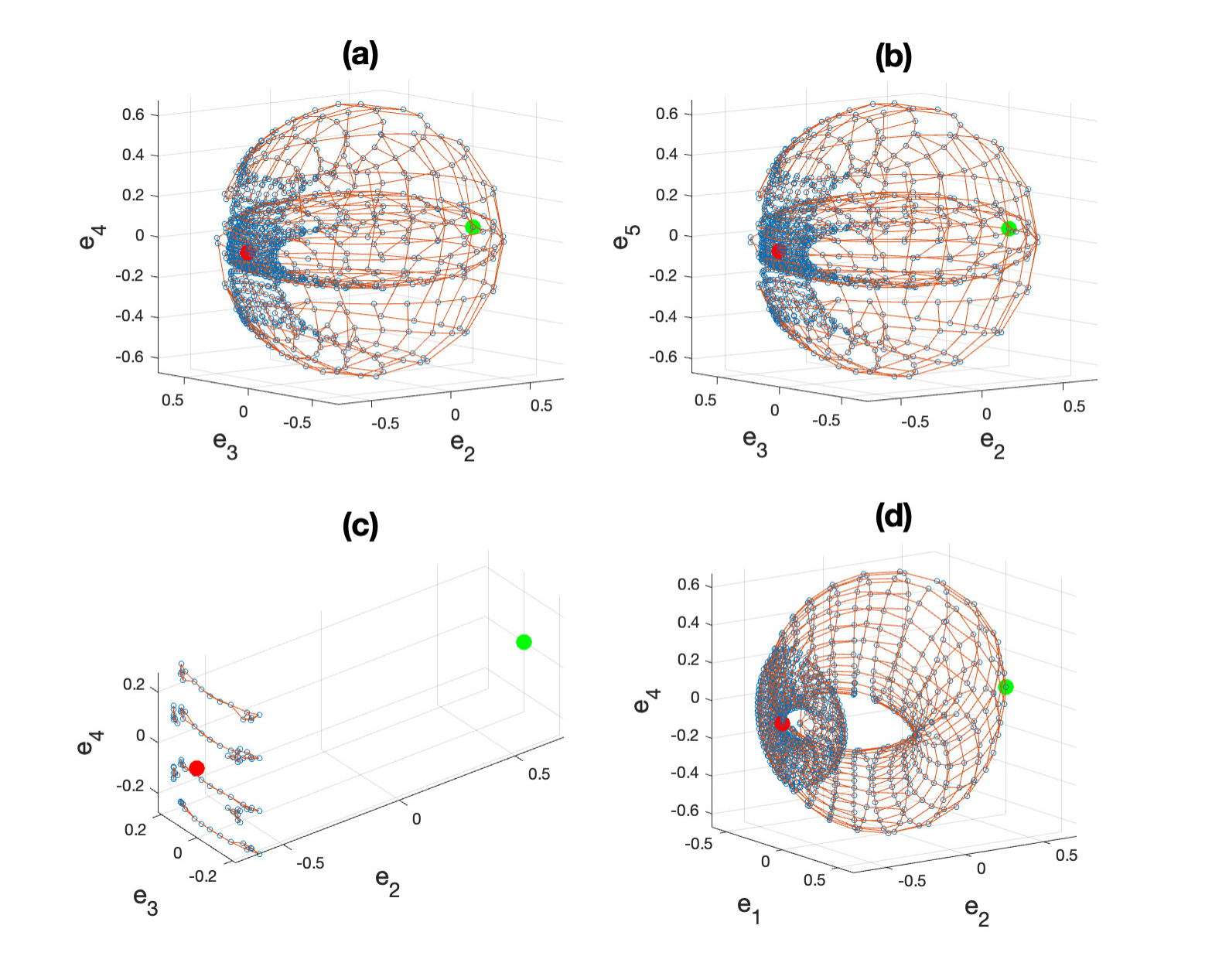}
    \caption{Manifold in five dimensions for the HOTI with $m_0=2$. The manifold projections into a three-dimensional space when one of the Cartesian coordinates is fixed: (a) and (b) $x_1=0$, (c) $x_1=0.05$, and (d) $x_3=0$. The red (green) solid circle is for states at $\Gamma$, $M$, and $Z (A)$, respectively.}
    \label{fig:manifold_HOTI}
\end{figure}

The phase diagram of the model indicates that when $1<\abs{m_0}<3$ and $\Delta_3=0$ the system is topological; otherwise it is non-topological. 
The phase diagram is reflected in the quantum parity in Fig.~\ref{fig:HOTI}: when the number of $d=2$ pair is odd, the system is topological. In a topological system, the maximum quantum disparity is $d=2$. However, unlike Chern insulators or Hopf insulators, only a few pairs of momenta exhibit $d=2$. This indicates that the manifold is not an ideal round sphere. Using the CMS method, we determine that the manifold's dimension for the HOTI ($m_0=2$) is five, necessitating consideration of projections. However, we find that when the data are projected into a three-dimensional space, the ``hole" will be filled. To reveal the ``hole", we show the projections of submanifolds on some four-dimensional plane. Figure~\ref{fig:manifold_HOTI} shows the projections of specified points where one coordinate is fixed: (a) and (b) $x_1=0$ (the first coordinate by our CMS method), (c) $x_1=0.05$, and (d) $x_3=0$. In Figs. \ref{fig:manifold_HOTI}(a) and (b), the projections form two concentric $S^2$'s (one situated in the exterior and the other compressed and positioned in the interior). The exterior points in (a) and interior in (b) correspond to states in the $k_x - k_y=0$ plane, while the opposite points correspond to states in the $k_x + k_y=0$ plane. When the $x_1$ coordinate is moved away from $0$, the $S^2$'s disappear as shown at $x_1=0.05$ in Fig. \ref{fig:manifold_HOTI}(b), excluding the possibility of a higher-dimensional sphere and indicating that the $x_1$ direction is topologically deformation retractable. In other cuts, for example at $x_3=0$ in Fig. \ref{fig:manifold_HOTI}(d), we do not find other sphere $S^n$ ($n\ge 2$). We conclude that the (covering space) manifold is homotopic to an $S^2$ and confirm that the axion insulator performs a $T^3$ to $S^2$ mapping (winding number 2).

\section{Conclusion} \label{sec:conclusion}

In this paper, we define a gauge-invariant distance between two quantum states. For a multilevel system, this distance can be conceptualized as a hyperplane. The relevant angles between two hyperplanes, known as principal angles, are determined within an appropriately chosen coordinate system. The cosines of these principal angles correspond to the singular values of the inner-product matrix of the two multilevel states. These principal angles provide insight into the quantum disparity, which quantifies the number of bases in which the two states differ. When the multilevel states form a multiband system over crystal momentum, this quantum disparity is related to a topological invariant. We demonstrate this relationship through examples involving quantum distance and disparity in various topological systems, including Chern insulators, Hopf insulators, and axion insulators.

To better visualize the manifold structure, we apply the multidimensional scaling method, a linear algebraic technique, to reconstruct the manifold based on the quantum distance. This method determines the minimal dimension of a Euclidean space required to embed the abstract manifold. The reconstruction can reveal nontrivial topology through the presence of ``holes" in the manifold. We anticipate that this method can be applied to other quantum systems to quantify their geometric information.

\section{Acknowledgements}
The research work of DG is supported by the National Science and Technology Council  of Taiwan with the Young Scholar Columbus Fellowship grant 113-2636-M-110-006. 
S.M.H. was supported by the NSTC-AFOSR Taiwan program on Topological and Nanostructured Materials, Grant No. 110-2124-M-110-002-MY3.

\appendix
\section{vectorization and triangle inequality} \label{App_A}
To have a sensible distance function $d(x,y)$ that measures the distance between each pair of elements $(x,y)$ in a manifold, certain properties or axioms must hold: (i) $d(x,x)=0$, (ii) $d(x,y)=d(y,x)$, and (iii) the triangle inequality $d(x,y) \le d(x,z) + d(z,y)$ for any elements $x,y,z$ in the manifold. These axioms imply non-negativity $d(x,y)\ge 0, \forall x,y$. While the first two properties are straightforward to demonstrate, the triangle inequality may not be as obvious. The quantum distance (or Hilbert-Schmidt distance) satisfies all three properties. The first two are clear, and the triangle inequality will be proved in this Appendix.

First, we show that the Hilbert-Schmidt distance is a valid distance function. Consider two quantum systems, $\Psi$ and $\Phi$, which are $k$-dimensional linear subspaces of an $n$-dimensional vector space. The Hilbert-Schmidt distance between them is defined as
\begin{equation}
    D_{\text{HS}}^{2}(\Psi,\Phi) = \frac{1}{2} \tr \left\{\left( P_{\Psi} - P_{\Phi} \right)^2 \right\},
\end{equation}
where $P_{\Psi},P_{\Phi}$ are projectors given by $P_{\Psi}=\sum_{i=1}^{k} \dyad{\psi_i}$ and $P_{\Phi}=\sum_{i=1}^{k} \dyad{\phi_i}$. The kets in each projector are orthonormal, so that $\tr P_{\Psi}^2= \tr P_{\Phi}^2 = k$, and hence $D_{\text{HS}}^{2}(\Psi,\Phi) = k - \tr \left(P_{\Psi} P_{\Phi} \right)$. Although the projectors are symmetric matrices, they can also be regarded as vectors. Since $\tr \left(P_{\Psi} P_{\Phi} \right)=\sum_{i,j} \left(P_{\Psi} \right) _{ij} \left(P_{\Phi} \right) _{ji} = \sum_{i,j} \left(P_{\Psi} \right) _{ij} \left(P_{\Phi} \right) _{ij}^* $, we can rearrange the pair of matrix indices $ij$ into a single vector index $\alpha$ as $ P_{ij} \rightarrow \widetilde{\vb{V}}_{\alpha}$, which norm is $k$. Although the vectors might be complex, the realness of $\tr \left(P_{\Psi} P_{\Phi} \right)= \widetilde{\vb{V}}_{\Psi} \vdot \widetilde{\vb{V}}_{\Phi}^* =  \widetilde{\vb{V}}_{\Psi}^* \vdot \widetilde{\vb{V}}_{\Phi}$ allows us to write it as an inner product of two real vectors, where each vector with double the size combines the real and imaginary parts of the corresponding $\widetilde{\vb{V}}$. The vectorization is suggested to be $ P \rightarrow \sqrt{2} \vb{V}_{\alpha}$, and therefore the Hilbert-Schmidt distance is geometrically the distance between two real vectors: $D_{\text{HS}}(\Psi,\Phi) = \| \vb{V}_{\Psi} - \vb{V}_{\Phi} \|$. With this, it becomes easier to prove the triangle inequality in a Euclidean space. By virtue of the Cauchy–Schwarz inequality, $\| \vb{a}\| \| \vb{b}\| \ge \abs{\vb{a \cdot b}}$, we have $\|\vb{a}\|+\|\vb{b}\| \ge \|\vb{a+b} \|$. Taking $\vb{a}=\vb{V}_{\Psi_1}-\vb{V}_{\Psi_3}$ and $\vb{b}=\vb{V}_{\Psi_3}-\vb{V}_{\Psi_2}$ into the above inequality, we 
can understand the triangle inequality for the Hilbert-Schmidt distances:
\begin{equation}
    D_{\text{HS}}(\Psi_1,\Psi_3) + D_{\text{HS}}(\Psi_3,\Psi_2) \ge D_{\text{HS}}(\Psi_1,\Psi_2). \label{DHS}
\end{equation}
All the properties shown above confirm a metric structure for quantum systems.

\section{Classical multidimensional scaling (CMS)} \label{App_B}
Multidimensional scaling (MDS) is an algorithm designed to to unravel the underlying manifold with a given table of distances between pairs of entities. The entities, which are Bloch states in our discussion, are viewed as nodes distributed in an abstract space. Their positions are determined according to the distance table. The reconstructed manifold will appear as an imaginary smooth surface that contains all nodes, resembling a map of entities.

Classical MDS assumes Euclidean distances. 
Given a distance matrix $D$ for the pairwise distances among $N$ entities, we aim to assign each entity a coordinate $\vb{X}_i$ for $i=1,\hdots, N$, such that $D_{ij} \approx \norm{\vb{X}_i - \vb{X}_j}$ as closely as possible. Assume that the vectors $\vb{X}_i $ have $n$ components as a row matrix, and let the collection of vectors form the $N$-by-$n$ matrix $X$: 
\begin{equation}
    X = 
    \begin{pmatrix}
    \vb{X}_1 \\
    \vb{X}_2 \\
    \hdots \\
    \vb{X}_N
\end{pmatrix}.
\end{equation}
The coordinates are encoded in the distance matrix $D$ since
\begin{equation}
    D_{ij} = \norm{\vb{X}_i - \vb{X}_j} = \left\{ \sum_{k=1}^{n} (X_{ik}-X_{jk})^2\right\}^{1/2}.
\end{equation}
From the squared distance matrix $A$,
\begin{equation}
    A_{ij}:=D_{ij}^2 =\norm{\vb{X}_i}^2 + \norm{\vb{X}_j}^2 - 2\vb{X}_i \cdot \vb{X}_j, \label{D2}
\end{equation}
we have the inner product of coordinate vectors $\vb{X}_i \cdot \vb{X}_j$. The Gram matrix, $B=XX^T$, is given by $B_{ij} = \vb{X}_i \cdot \vb{X}_j = -\frac{1}{2} \left( A_{ij} -\norm{\vb{X}_i}^2-\norm{\vb{X}_j}^2 \right)$.

The solution for $X $ is not unique. 
To ensure a centered configuration, we impose the condition $\frac{1}{N}\sum_{i=1}^{N} \vb{X}_i = \vb{0}$, is imposed. By summing over $j$ in Eq. (\ref{D2}), we find the norm of the vector 
\begin{equation}
    \norm{\vb{X}_i}^2 =\frac{1}{N} \sum_{j} A_{ij} - \frac{1}{N} \sum_{j}  \norm{\vb{X}_j}^2.
\end{equation}
Summing over $i $ as well, we have
\begin{equation}
    \frac{1}{N} \sum_{i} \norm{\vb{X}_i}^2 = \frac{1}{2N^2} \sum_{i,j} A_{ij}.
\end{equation}
After some derivations, the $B$ matrix is produced from $A$ as 
\begin{equation}
    B_{ij} = -\frac{1}{2} \left( A_{ij} -\frac{1}{N} \sum_{i} A_{ij} -\frac{1}{N} \sum_{j} A_{ij} + \frac{1}{N^2} \sum_{i,j} A_{ij}\right),
\end{equation}
or in matrix form
\begin{equation}
    B = -\frac{1}{2} C A C,
\end{equation}
where the centering matrix is $C=I_{N}-\frac{1}{N}J_{N}$ with $I_N$ being the identity matrix and $J_N$ an $N\times N$ matrix of all ones. 

By diagonalizing $B$, we have $B=V\Lambda V^T$. The eigenvalue matrix $\Lambda$ contains $N$ eigenvalues $\lambda_i$ and the eigenvector matrix $V$ contains $N$ eigenvectors. Typically, $\rank B <N$, and we may discard eigenvectors associated with small eigenvalues. Thus, the eigenvalue and eigenvector matrices are reduced in their dimensions: $\Lambda \rightarrow \Lambda_r = \text{diag} \{\lambda_1, \hdots,\lambda_n \}$ and $V \rightarrow V_r = \left(\vec{v}_1, \hdots,\vec{v}_n \right)$, where $\vec{v}_i$ are column vectors. Assuming $B$ is positive-semidefinite ($\lambda_i\ge 0$), we define the matrix $Y=V_r \Lambda_r^{1/2}$, giving $B = YY^T$. The rows of $Y$, $\vb{Y}_i$, are the reconstructed coordinates (also called principal coordinates) of the entities:
\begin{equation}
    Y = 
    \begin{pmatrix}
    \vb{Y}_1 \\
    \vb{Y}_2 \\
    \hdots \\
    \vb{Y}_n
\end{pmatrix} =
\begin{pmatrix}
    \sqrt{\lambda_1} \vec{v}_1,
    \hdots, 
    \sqrt{\lambda_n} \vec{v}_n
\end{pmatrix} ^T.
\end{equation}

\bibliographystyle{apsrev4-1}
\bibliography{qdistance}

\end{document}